# A General Equilibrium Theorem for the Economy of Giving

*W.P. Weijland*

Informatics Institute, Faculty of Science, University of Amsterdam, Science Park 904, 1098 XH Amsterdam, the Netherlands

November 2014

**Abstract.** In [1] we presented a model for transactions when goods are given away in the expectation of a later settlement. In settings where people keep track of their social accounts we were able to redefine concepts like account balance, yield curve and the law of diminishing returns. In this paper we establish a general equilibrium theorem, conjectured in [1], by developing sufficient conditions for any instance of the standard model (or Gift Economy Model) to have a unique equilibrium. The convergence to that equilibrium is exponential and for each pair of entities P and Q the total sum of yields from all mutual transactions is equal to zero.

### I.    INTRODUCTION

This article serves as the mathematical underpinning of some of the claims put forward in [1]. It may be regarded as a record of proof for the assertions in that paper serving future work that may be based on it.

Future work may lead into several directions. In [4] it was shown that the Economy of Giving model can be used to optimize between leechers and seeders in a peer-to-peer file sharing network such as BitTorrent. In [6] ways are suggested how a mutual account balance system such as the Economy of Giving model can be scaled up from small communities to large populations. Other directions of application are those in the realm of microeconomics and game theory. The Kiyotaki-Wright model ([5]) for example, that proves the emergence of money-like commodities ([7]) in moneyless exchange trade markets, provides the basis for a natural extension to 'gift commodities' in a small-size gift community.

Clearly, the mathematical work has not ended after this paper. The central convergence theorem presented here may be used to identify other convergent instances of the model – e.g. other than the one with 'linear yield curves' presented in [1] – emerging from concrete applications in practical setting settings.

This paper does not address the concept of the 'highest yield rule' as conceived in [1]. Attempts to formulate such result gave the impression that a general equilibrium theorem may be possible but is far more complex than the one for the basic model presented in this paper, mainly due to discontinuities arising at the intersection points of the yield curves.

We start off with a brief introduction along the lines as presented in [1].

## II    THE STANDARD MODEL[1]

In order to formulate our key results we need some concepts and terminology first.

DEFINITION II.1    A *multiset* is a non-ordered, final set with multiple occurrences of elements. For example, we write M = [1,1,2,5,7,4,7,7,8] for a nine-element multiset of natural numbers 1, 2, 4, 5, 7 and 8 with occurrences 2, 1, 1, 1, 3, 1 respectively[2]. A *multiset over a set A* is a multiset containing only (multiple) elements from set A.

If M is a multiset over A then all domain elements in A not occurring in M are said to have occurrence zero in M. We write a∈M if the occurrence of a in M is greater than zero.

A multiset K is a *multi-subset* of multiset L, notation: K⊆L, if all elements occur at most as many times in K as they do in L. A *multipair* is a multiset with just two elements.

For multisets K and L we write K ∪ L for the multiset containing only elements from K or L and in which all elements occur precisely as many times as the sum of their occurrences in K and L respectively.

DEFINITION II.2    An *entity* is a person, computer, group, network or any other subject able to interact with other entities through the transfer of goods, where a *good* is any type of product, service or favor with a (positive) economic value. All entities under consideration together are referred to as the *community*.

DEFINITION II.3    The act of offering a good *a* to all entities by entity P is called the *supply of a by P*. Notation: P —$a$→ . Entity P is called a *supplier* of *a*.
The act of accepting a good *a* from any entity by Q is called a *demand of a by Q*, written as: —$a$→Q. Entity Q is called a *recipient* of *a*.

The supply P —$a$→ and demand —$a$→Q together may lead to a *transaction of a from P to Q*, notation: $t$ = P —$a$→ Q. Such transaction $t$ consists of a multipair [P —$a$→, —$a$→Q] of the supply and the demand of the same product *a*. A transaction is said to be a *transaction between P and Q* if it is a transaction from P to Q or a transaction from from Q to P. P and Q are *involved* in transaction $t$.

DEFINITION II.4    The *Supply-Demand Space* SDS is defined as:
SDS = { P —$a$→, —$b$→Q : for all entities P, Q and all goods *a*, *b*}.
A *state* is a multi-subset over SDS. Transaction $t$ = P —$a$→ Q is called *admissible* in state S if P —$a$→ ∈S and —$a$→Q ∈S.

A multiset of transactions T = [$t_1$, $t_2$, $t_3$, ..., $t_k$] is *admissible* in state S if $\cup_{i\in \mathbf{N}} t_i \subseteq$ S.
In other words: S caters for all supply and demand needed to enable all transactions in T.

DEFINITION II.5    The *standard model* consists of a sequence of pairs $(S_i, T_i)_{i\in \mathbf{N}}$ of states $S_1, S_2, S_3, ...$ and multisets of transactions $T_1, T_2, T_3, ...$ such that each $T_i$ is admissible in $S_i$.

Where $S_i$ is the space of all supply and demand at point i, $T_i$ can be viewed as the set of transactions actually realized at that point.

---

[1] This section is taken from [1].
[2] More precisely, we note that a multiset is a set (the universe) together with a mapping from its elements to $\mathbf{N_0}$ (respective occurences).

DEFINITION II.6  For each entity P and at each point i we assume the existence of a *yield function* $\cdot^P$ mapping each possible transaction $t$ from $T_i$ to a real number such that:

(i)  If P is not involved in $t$ then $t^P = 0$
(ii) If $t = P \xrightarrow{a} Q$ for some Q and $a$ then $t^P \geq 0$
(iii) If $t = Q \xrightarrow{a} P$ for some Q and $a$ then $t^P \leq 0$.

One may look at $t^P$ as the value attached by P to a certain transaction. Note that we have such a yield function in each point i so that P may value the same transaction differently over each point. However, we will assume that yield functions do not vary over the state index i.

The yield functions of P and Q are *competible* if: $t^P + t^Q = 0$, for all transactions $t$ between P and Q. This implies that P and Q share the same view on the value of each transaction that they enter into. The notion of competibility also allows us to define a relative account balance between P and Q:

DEFINITION II.7  At point i the *account balance* $x = A_{i,P,Q}$ of P with respect to Q is defined as:

(i)  $A_{0,P,Q} = 0$
(ii) $A_{i+1,P,Q} = A_{i,P,Q} + \sum_t t^P$, where $t$ ranges over all transactions in $T_i$.

Note that if the yield function is competible then $A_{i,P,Q} + A_{i,Q,P} = 0$ voor all i.

We assume the yield $t^P$ is a function of the account balance $x = A_{P,Q}$ between P and Q.

DEFINITION II.8  Given entities P and Q and transaction $t = P \xrightarrow{a} Q$ the *yield curve* of P is the function $P(x)$ that for every account balance $x = A_{P,Q}$ produces the yield $t^P$. We assume that for every (P, Q, $t$) there is one unique such function.

DEFINITION II.9  The *Law of Diminishing Returns* expresses that for all (P, Q, $t$) the corresponding yield curve is monotonically non-increasing.

Suppose $t = P \xrightarrow{a} Q$. Under the Law of Dimishing Returns the yield curve may look as follows:

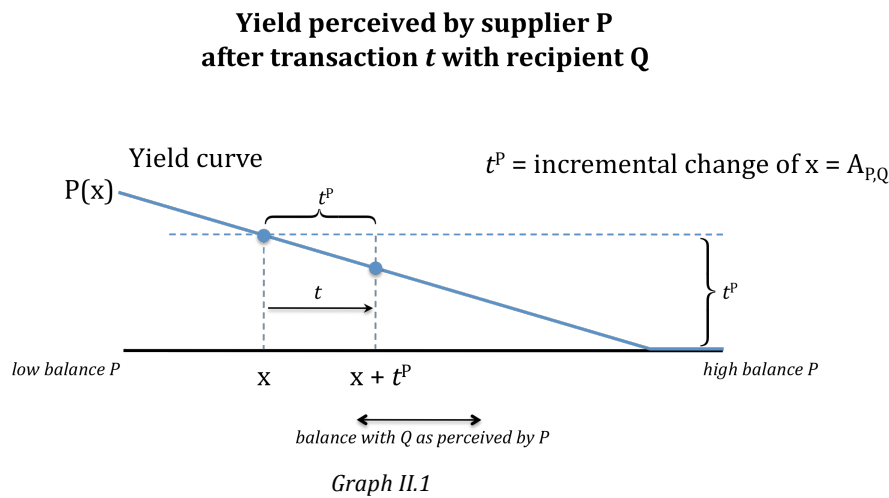

Graph II.1

In this graph we have assumed the yield curve to be linear. Notice that the dotted lines together enclose a square shape representing the step $t$ down the curve. More steps

down the curve lead to:

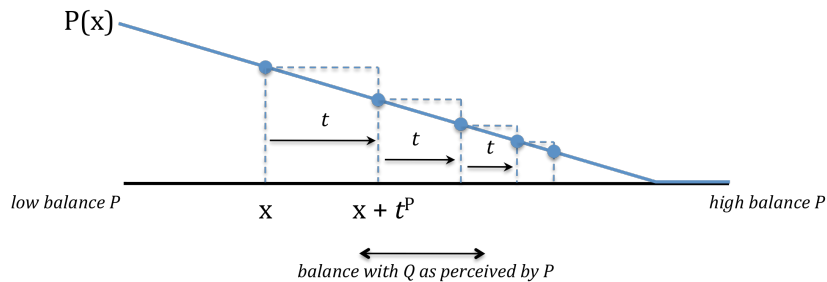

Graph II.2

obviously converging to the fixed point where P(x) = 0. In [1] we have shown that this convergence is exponential. Note that the yield of each next transaction $t$ is smaller than any of the previous ones. This reflects the law of diminishing returns ("law of decreasing yield").

We assume $t^P$ and $t^Q$ to be compatible so that whenever P's balance with Q is x then Q's balance with P is –x. Plotted in a *transaction diagram*:

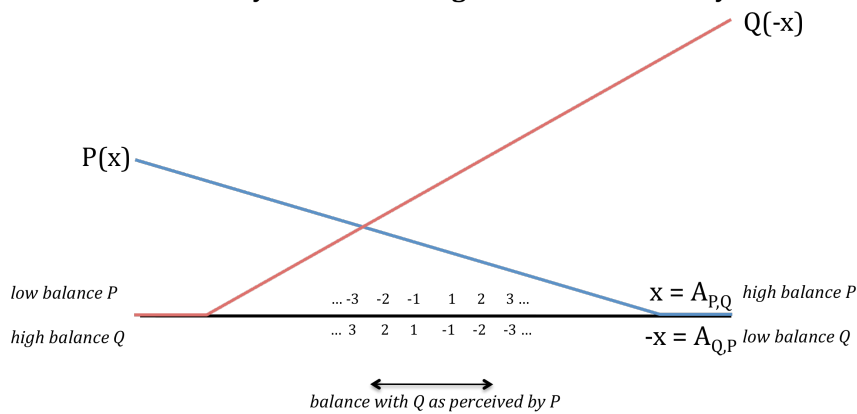

Graph II.3

Projecting Q as a mirror image in the coordinate system reflects that each increase of the balance of P implies an equal decrease of the balance of Q, and vice versa. So this way they share the same scale.

Suppose the balance between P and Q is $x_0$. Below is the transaction diagram when P starts with $t$ = P $\xrightarrow{a}$ Q and then Q follows with $t'$ = Q $\xrightarrow{b}$ P.

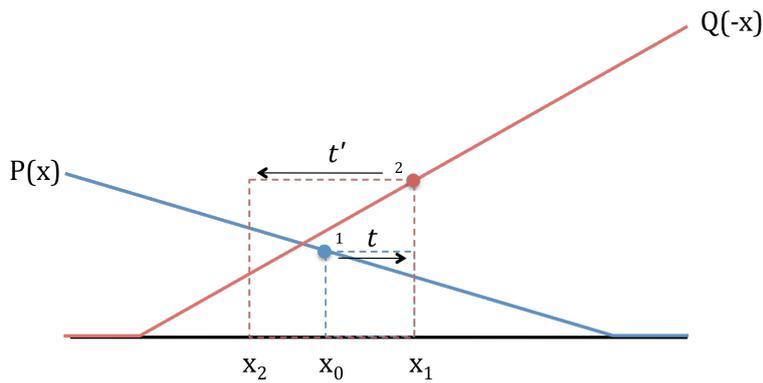

**Two transaction steps by P and Q respectively resulting in a balance change from $x_0$ to $x_1$ and then $x_2$**

*Graph II.4*

The transactions $t$ and $t'$ may also occur simultaneously as in graph II.5:

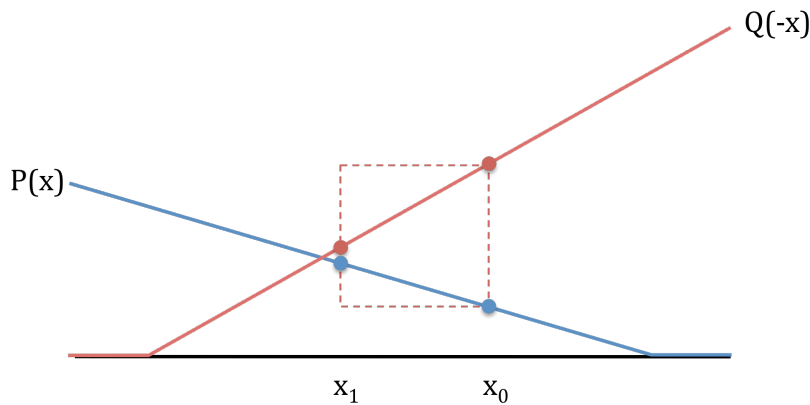

**Two simultaneous transactions by P and Q respectively resulting in a change of balance from $x_0$ to $x_1$**

*Graph II.5*

The square shape results from the subtraction $Q(–x_0) – P(x_0)$ and the completion of the transaction square to find the new balance $x_1$ between P and Q.

This paper is about equilibriums that may occur in transaction diagrams such as above. For example: it is not difficult to see that in graph II.5 the repeated simultaneous transactions $t$ and $t'$ converge to the intersection point of both yield curves. At this intersection point the yields of $t$ and $t'$ are equal and the simultaneous exchange of goods does not change the balance any longer.

When the transactions $t$ and $t'$ are not simultaneous we find an equilibrium of a different shape than just the point of intersection:

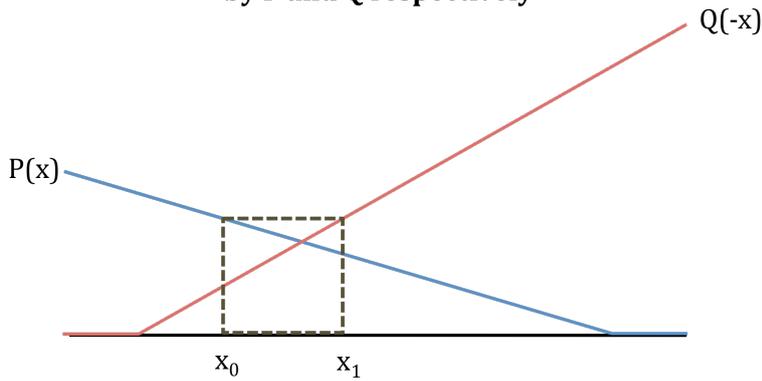

Graph II.6

When P has balance $x_0$ with Q then the transaction $t = P \xrightarrow{a} Q$ results in the new balance $x_1$. If then Q follows with a transaction $t' = Q \xrightarrow{b} P$ then the balance returns to $x_0$. In [1] we found that from any initial starting point the alternating transaction process converges to this 2-fold equilibrium. Note that the formal representation of this process in the standard model reads as follows:

$S_i \quad := [P \xrightarrow{a}, Q \xrightarrow{b}, \xrightarrow{b} P, \xrightarrow{a} Q]$
$T_{2i-1} := [P \xrightarrow{a} Q]$
$T_{2i} \quad := [Q \xrightarrow{b} P]$
for all $i \in \mathbf{N}$.

However, the standard model allows for much more complex equilibriums. For example:

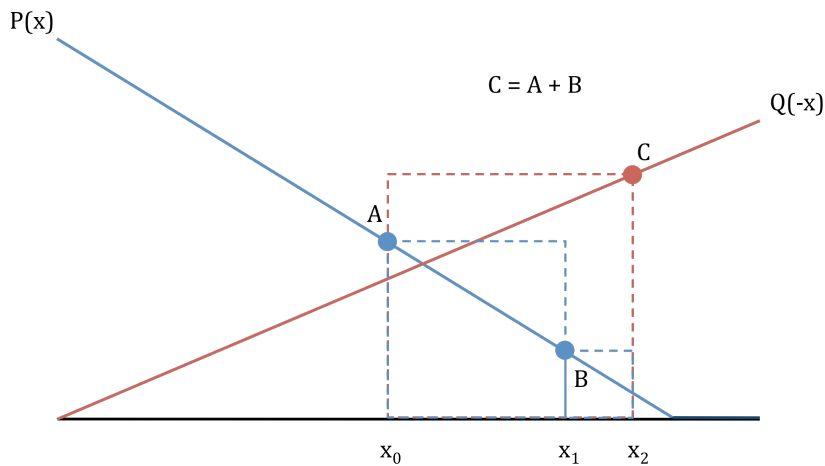

Graph II.7

This graph shows the 3-fold equilibrium resulting from:

$T_{3i-2} := [P \xrightarrow{a} Q]$
$T_{3i-1} := [P \xrightarrow{a} Q]$
$T_{3i} := [Q \xrightarrow{b} P]$
for all i∈**N**.

In words: 'after two times $P \xrightarrow{a} Q$ follows one time $Q \xrightarrow{b} P$'. Note that in this equilibrium we have C = A + B, where in equations A, B and C stand for the respective yields at points A, B and C. To see this, note that A, B and C each are at a vertex of a square.

A variant of this looks as:

**Equilibrium of three alternating transactions:
one by P (at $x_0$) and two by P and Q simultaneously (at $x_1$)**

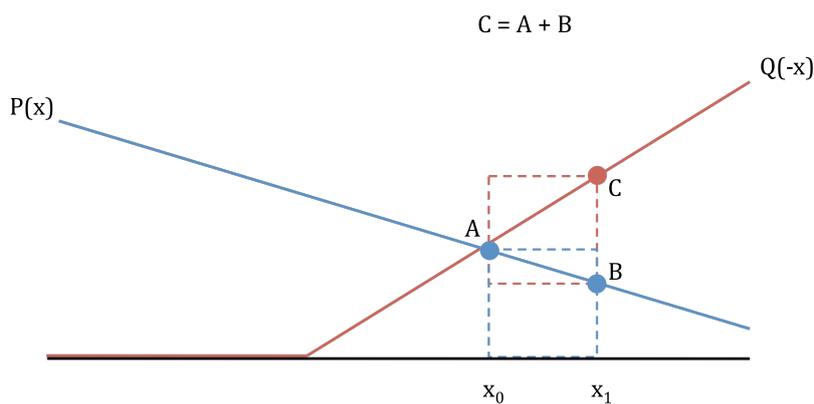

*Graph II.8*

Graph II.8 shows the 2-fold equilibrium of:

$T_{2i-1} := [P \xrightarrow{a} Q]$
$T_{2i} := [P \xrightarrow{a} Q, Q \xrightarrow{b} P]$
for all i∈**N**.

Again we find C = A + B since C – B = ($x_1 - x_0$) = A (note that A is the vertex of a square that is congruent to the square with vertices B and C).

Some equilibriums can look pretty complex such as:

**Equilibrium of simultaneous transactions from P and R at $x_0$ and $x_2$
and one separate transaction by R at $x_1$**

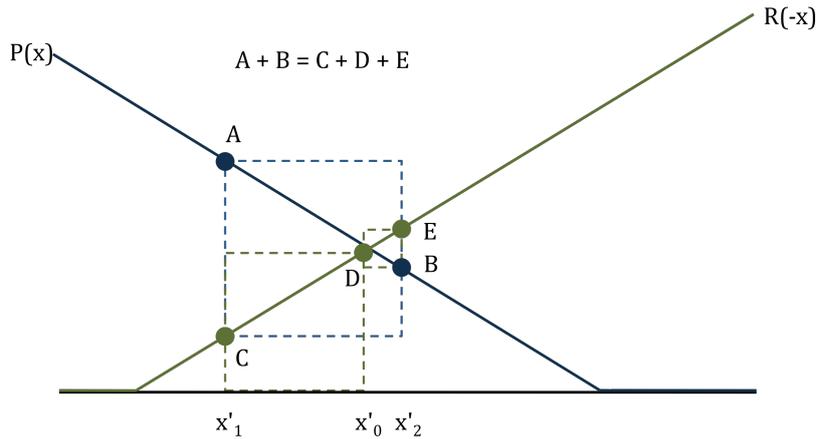

Graph II.9

This 3-fold equilibrium results from:

$T''_{3i-2} := [P \xrightarrow{a} R, R \xrightarrow{c} P]$
$T''_{3i-1} := [P \xrightarrow{a} R, R \xrightarrow{c} P]$
$T''_{3i} \;\;\; := [R \xrightarrow{c} P]$
for all $i \in \mathbf{N}$.

This time P and R exchange goods *a* and *c* except for every third time when P fails to supply good *a*.

It is easy to check that $A - C = D + (E - B)$ and hence $A + B = C + D + E$. That is: all transactions by P add up to the sum of all transactions by R. In fact this turns out to be a general 'zero sum' property of any equilibrium.

The question at hand in this paper is: does such equilibrium always exist? And is it unique? The answer turns out to be affirmative, provided some preconditions are met. We will formulate two convergence theorems and construct their proofs.

### III    PREPARATIONS
We need some definitions first before we can formulate the final result.

DEFINITION III.1    In a complete metric space $(X, d)$ the mapping $C: X \to X$ is called
  i. *non-expanding* if for all $x,y \in X$:
     $d(C(x),C(y)) \leq d(x,y)$
  ii. a *contraction* if there exists $0 \leq q < 1$ such that for all $x,y \in X$:
     $d(C(x),C(y)) \leq q \cdot d(x,y)$.

A central result in fixed point theory for metric spaces is Banach's Contraction Mapping Theorem which states that every contraction mapping C in a complete metric space $(X,d)$ has a unique fixed point, i.e: a unique point x with $C(x) = x$ (see [2]). Moreover, this point can be found by iterating C infinitely many times from any arbitray starting point $x_0 \in X$. Hence for any $x_0 \in X$: $C^k(x_0) \to x$ when $k \to \infty$.

Note that every contraction is non-expanding and every non-expanding mapping is continuous. Further we have:

LEMMA III.2    (composition)
  i.  The composition of two non-expanding mappings is again non-expanding.
  ii. The composition of a non-expanding mapping and a contraction is a contraction.

The proof is straightforward and left to the reader.

DEFINITION III.3    A function F: $\mathbf{R} \to \mathbf{R}$ is called *uniformly monotonous* if there exists some r>0 such that for all x,y∈$\mathbf{R}$:  $r \cdot |x - y| \leq |F(x) - F(y)|$.

The definition implies that F de(in)creases at least with speed r. Note the if F is uniformly monotonous and continuous then it is either monotonically increasing or monotonically decreasing.

DEFINITION III.4    A multiset of transactions T is called *basic* if for every pair of entities (P, Q) there exists at most one transaction from P to Q in T. An instance $(S_i, T_i)_{i \in \mathbf{N}}$ of the standard model is *basic* if $T_i$ is a basic multiset of transactions for all i∈$\mathbf{N}$.

The notion of a basic model implies that P can only give away to Q one good at the time. If at a certain instance i we would allow P to give away more than one good to Q then we would need to make assumptions on the yield curve of such combination of goods and hence on the change of the account balance when such combination of goods is given away.

DEFINITION III.5    For every pair of entities P, Q and transaction $t$ from P to Q we assume we have a unique yield curve $P_{Q,t}$: $\mathbf{R} \to \mathbf{R}^+$ (often abbreviated 'P' when it is clear from the context that it concerns Q and $t$), where $\mathbf{R}$ are the real numbers and $\mathbf{R}^+$ are the non-negative real numbers (hence including zero).

ASSUMPTION III.6    For any (Q, $t$) we assume the corresponding yield curve $P_{Q,t}$ to be:
  1. Monotonically non-increasing
  2. Non-expanding.

The assumption is in line with [1] where we assumed yield curves to be linear[3] and non-increasing with coefficient -1<a≤0 (hence non-expanding). This time we allow yield curves to be non-linear provided they are a non-expanding.

In order to formalize the notion of account balance in the standard model we need the following definition:

DEFINITION III.7    (Account Operator)  Let $(S_i, T_i)_{i \in \mathbf{N}}$ be a basic instance of the standard model. Let P and Q be entities in $(S_i, T_i)_{i \in \mathbf{N}}$. Since $(S_i, T_i)_{i \in \mathbf{N}}$ is basic each $T_i$ has at most one transaction from P to Q, denoted by $t_i$, and at most one transaction from Q to P, denoted by $t'_i$. Let $P_i$ and $Q_i$ be the yield curves corresponding to the transactions $t_i$ and $t'_i$ respectively. If $T_i$ has no transaction from P to Q then we assume $P_i(x) = 0$ for all x∈$\mathbf{R}$. and if $T_i$ has no transactions from Q to P then we assume $Q_i(x) = 0$ for all x∈$\mathbf{R}$. Note that $P_i$ and $Q_i$ are non-expanding for all i∈$\mathbf{N}$.

---
[3] To be precise: linear until zero from which they remain flat.

Now, for i∈**N** and x∈**R** define the *account operator* $A_i$: **R** → **R** as follows:

$A_i(x) := x + P_i(x) - Q_i(-x)$.

If $A_i(x) = x$ for all x (hence both $P_i$ and $Q_i$ are flat at zero) then $A_i$ is said to be *trivial*.

Let $x_0 \in$ **R** be the *initial balance* between P and Q and for i∈**N** define $x_i := A_i(x_{i-1})$. Then we refer to $(x_i)_{i \in \mathbf{N}}$ as *the balance sequence with initial balance* $x_0$, or simply *a balance sequence* when $x_0$ is taken as a variable. This definition formalizes the construction that we demonstrated in the transaction diagrams in the previous section II.

DEFINITION III.8    A sequence $(q_i)_{i \in \mathbf{N}}$ of objects $q_i$ is *cyclical* if for some *k* and all i: $q_{i+k} = q_i$. The smallest such number *k* is the *order* of recurrence of $(q_i)_{i \in \mathbf{N}}$.

A sequence is *recurrent* if it is cyclical from some point. Recurrency is comparable to the nature of rationals (with recurrent decimals) or to solutions of finite recursive specifications (for example see: [3]).

DEFINITION III.9    If a balance sequence is cyclical it is called an *equilibrium*. If the order is *k*, then such balance sequence is called a *k-fold equilibrium*.

DEFINITION III.10    An instance $(S_i, T_i)_{i \in \mathbf{N}}$ of the standard model is *cyclical* if $(T_i)_{i \in \mathbf{N}}$ is.

If $(T_i)_{i \in \mathbf{N}}$ is cyclical and basic then it follows from definition III.7 that also $A_i$ is cyclical with the same order. Now suppose $(A_i)_{i \in \mathbf{N}}$ has order *k* then $A_i = A_{i+k}$ for all i. In the balance sequence $x_i = A_i(x_{i-1})$ we have:

$x_{i+k} := A_{i+k}...A_{i+1}(x_i)$, for all i.

For given *k* define $C_i$: **R** → **R** such that:

$C_i := A_{i+k}...A_{i+1}$

Then $x_{i+k} := C_i(x_i)$ and $C_{i+k} = C_i$ for all i, since $(A_i)_{i \in \mathbf{N}}$ is cyclical with order *k*. Finally, define $C$: **R**$^k$ → **R**$^k$ as follows:

$$C := \begin{pmatrix} C_1 \\ C_2 \\ ... \\ C_k \end{pmatrix}.$$

Transforming $(x_i)_{i \in \mathbf{N}}$ into vector notation we define:

$$\underline{x}_i := \begin{pmatrix} x_{ik+1} \\ x_{ik+2} \\ ... \\ x_{(i+1)k} \end{pmatrix} \quad \text{for all i} \in \mathbf{N}_0.$$

So starting from $x_0$ the vector sequence $(\underline{x}_i)_{i \in \mathbf{N}}$ takes every next *k* values in $(x_i)_{i \in \mathbf{N}}$ as a new vector. Then it easily follows from the construction of $C$ that for all i:

$\underline{x}_{i+1} = C(\underline{x}_i)$.

## IV STATIC CONVERGENCE

The convergence theorem is founded on the notion that if the operator $C$ would be a contraction in $\mathbf{R}^k$ then for any starting point $\underline{x}_0$ the iteration $CC...C(\underline{x}_0)$ converges to a unique fixed point of the contraction $C$. So the question is: when is $C$ a contraction? This question will be explored below.

LEMMA IV.1  Suppose $P(x)$ and $Q(x)$ are non-expanding and monotonically non-increasing then $L(x) := x + P(x) + Q(x)$ is non-expanding.

PROOF  Choose $x, y \in \mathbf{R}$. Without loss of generality we assume $x \leq y$. Since $x \leq y$ and P is non-increasing we have: $0 \leq (P(x) - P(y))$. Since P is non-expanding we also have: $|P(x) - P(y)| \leq |x - y| = (y - x)$. Hence: $0 \leq (P(x) - P(y)) \leq (y - x)$. Likewise we also have: $0 \leq (Q(x) - Q(y)) \leq (y - x)$ and adding up these inequalities we find:
$$0 \leq \{P(x) - P(y) + Q(x) - Q(y)\} \leq 2.(y - x).$$
Adding $(x - y)$ to all sides we find:
$$(x - y) \leq \{x - y + P(x) - P(y) + Q(x) - Q(y)\} \leq (y - x)$$
or:  $(x - y) \leq (L(x) - L(y)) \leq (y - x)$
Hence $|L(x) - L(y)| \leq |x - y|$ and L is non-expanding.  □

COROLLARY IV.2  Suppose $P(x)$ and $Q(x)$ are non-expanding and monotonically non-increasing then $A(x) := x + P(x) - Q(-x)$ is non-expanding.

PROOF  Since $Q(x)$ is non-expanding and monotonically non-increasing, so is $-Q(-x)$. Now use Lemma IV.1.  □

COROLLARY IV.3  If in definition III.7 the yield curves $P_i$ and $Q_i$ are non-expanding and non-increasing then $A_i$ is non-expanding for all i.

If all $A_i$ are non-expanding so is the composition $C_i$ of account operators and hence so is $C$ in $\mathbf{R}^k$ in case $(A_i)_{i \in \mathbf{N}}$ is cyclical with order $k$. However: being non-expanding is not sufficient for $C$ to have a fixed point. Even if we require all yield curves to be contractions we cannot prove that $C$ is a contraction as can be seen from the trivial case with one flat curve $P(x) = 0$ which has fixed points everywhere, or a flat constant curve $P(x) = 1$ which causes the account sequence to diverge to infinity.

LEMMA IV.4  Suppose P and Q are non-expanding and non-increasing. If P is uniformly monotonous and Q is a contraction then
$L(x) := x + P(x) + Q(x)$ is a contraction.

PROOF  Since P is uniformly monotonous and non-expanding (hence continuous) there exists r>0 such that for all $x, y \in \mathbf{R}$: $r.|(x - y)| \leq |P(x) - P(y)| \leq |x - y|$. Clearly $r \leq 1$. Suppose $r=1$ then we have $|P(x) - P(y)| = |x - y|$ and since P is continuous and non-increasing it has the form $P(x) = -x + c$ for some constant c. Hence $L(x) = c + Q(x)$ and since Q is a contraction, so is L.
So assume $0 < r < 1$. Choose $x, y \in \mathbf{R}$. Without loss of generality we assume $x \leq y$ and hence $P(y) \leq P(x)$, since P is non-increasing. Again we have:
  (1) $r.(y - x) \leq (P(x) - P(y)) \leq (y - x)$.
Since Q is a contraction we have: $|Q(x) - Q(y)| \leq q.|x - y|$ for all $x, y \in \mathbf{R}$. Since $x \leq y$ and Q is non-increasing we have $Q(y) \leq Q(x)$ and so:
  (2) $0 \leq (Q(x) - Q(y)) \leq q.(y - x)$.
Adding up (1) and (2) we find:

$$r.(y - x) \leq \{P(x) - P(y) + Q(x) - Q(y)\} \leq (1 + q).(y - x).$$

Adding $(x - y)$ to all sides we then have:

$$(1 - r).(x - y) \leq \{x - y + P(x) - P(y) + Q(x) - Q(y)\} \leq q.(x - y)$$

or: $\quad (1 - r).(x - y) \leq (L(x) - L(y)) \leq q.(x - y)$

so: $\quad |L(x) - L(y)| \leq MAX(|1 - r|, q).|x - y|$.

Since $0 < r < 1$ we have $0 < (1 - r) < 1$ and so: $0 < MAX(1 - r, q) < 1$. Hence L is a contraction. □

The intuition behind the lemma can be understood easier if we assume P and Q to be differentiable:

(i) Since P is uniformly monotonous and non-expanding we have for all x:
$0 < r \leq |P'(x)| \leq 1$;

(ii) Since Q is a contraction we have $0 \leq |Q'(x)| \leq q < 1$ for all x;

(iii) Since P and Q are non-increasing we have: $P'(x) \leq 0$ and $Q'(x) \leq 0$ for all x;

(iv) $L'(x) = 1 + P'(x) + Q'(x)$.

Now one can easily see that $1 - 1 - q \leq L'(x) \leq 1 - r - 0$ or: $-q \leq L'(x) \leq 1 - r$ for all x and therefore $|L'(x)| \leq MAX(1-r, q)$. Hence L is a contraction. Check that lemma IV.4 also holds if P is a uniformly monotonous contraction and Q is just non-expanding.

COROLLARY IV.5          Let $i \in \mathbf{N}$. If in definition III.7 both $P_i$ and $Q_i$ are non-expanding and monotonically non-increasing and:
- at least one out of $P_i$ and $Q_i$ is uniformly monotonous
- at least one out of $P_i$ and $Q_i$ is a contraction

then $A_i$ is a contraction.

PROOF      One can easily prove that:

(i) If $P_i(x)$ and $Q_i(x)$ are monotonically non-increasing, so are $-P_i(-x)$ and $-Q_i(-x)$.

(ii) If $P_i(x)$ is uniformly monotonous, so is $-P_i(-x)$.

(iii) If $P_i(x)$ is a contraction, so is $-P_i(-x)$.

Now use Lemma IV.4. □

THEOREM IV.6 (Convergence)     Let $(S_i, T_i)_{i \in \mathbf{N}}$ be a basic and cyclical instance of the standard model with order *k*. Let P and Q be entities in $(S_i, T_i)_{i \in \mathbf{N}}$ and let $A_i$, $P_i$ and $Q_i$ be as in definition III.7. Suppose we have some closed subset $\mathbf{I} \subseteq \mathbf{R}$.

Then if:
1. **I** is closed with respect to all operators $C_i$
2. All $P_i$ and $Q_i$ are monotonically non-increasing and non-expanding on **I**
3. For some $1 \leq j \leq k$:
   a. at least one out of $P_j$ and $Q_j$ is uniformly monotonous on **I**
   b. at least one out of $P_j$ and $Q_j$ is a contraction on **I**

then the operator $C: \mathbf{R}^k \rightarrow \mathbf{R}^k$ from section III has a *k*-fold equilibrium that is unique.

PROOF Since all $P_i$ and $Q_i$ are non-expanding and monotonically non-increasing on **I** we find with Corollary IV.3 that all $A_i$ are non-expanding on **I**. Let $j \in \mathbf{N}$ such that one out of $P_j$ and $Q_j$ is uniformly monotonous and also at least one out of $P_j$ and $Q_j$ is a contraction on **I** then it follows from Corollary IV.5 that $A_j$ is a contraction on **I** and by Lemma III.2.(ii) that $C_i$ is a contraction for all i. Hence $C$ is a contraction on $\mathbf{I}^k$ and by Banach's contraction theorem $C$ has a unique fixed point <u>u</u> representing a *k*-fold equilibrium. □

We can use theorem IV.6 to prove that all previous examples from the Graphs II.5-9 represent converging models. To see this, note that all yield functions in these examples:
- are non-expanding
- are monotically non-increasing
- are positive for x<0
- have an intersection point with the *x*-axis from where they are flat at zero.

Now suppose at least one $P_i$ is not flat at zero. We check the conditions of theorem IV.6.
Condition 1: Pick the yield function $P_s$ with the largest intersection point with the *x*-axis, let us say this is at point x=z. Clearly z≥0 and $P_i(z) = 0$ for all i. Define: **I** := (∞,z]. Since $P_i$ is a contraction for all i we have $|P_i(x) - P_i(z)| \leq |x - z|$ for all x≤z. Using that $P_i(z) = 0$ and x≤z we find $P_i(x) \leq (z - x)$ and so $x + P_i(x) \leq z$. Since all $Q_i$ are non-negative we also have $A_i(x) = x + P_i(x) - Q_i(-x) \leq z$ for all x≤z. Thus **I** is closed with respect to the operators $A_i$.
Condition 2: All $P_i$ and $Q_i$ are monotonically non-increasing contractions on **I**.
Condition 3: Clearly, $P_s$ is a non-negative, decreasing linear function on **I** with $P_s(z) = 0$ and $P_s(x) > 0$ for all x<z. It is easy to see that $P_s$ is a confined contraction on **I**.
Finally, use theorem IV.6 to conclude that from any starting point $x_0$ in **I** the cyclical sequence defined by $x_i := A_i(x_{i-1})$ has a unique *k*-fold equilibrium in **I**.

If also at least one $Q_i$ is not flat at zero one can prove in the same way that **I'** := [z', ∞) is closed with respect to all operators $A_i$, where z' is the smallest intersection point with the *x*-axis of all functions $Q_i$. Clearly z'≤0. Hence from any starting point $x_0$ in **I'** the cyclical sequence defined by $x_i := A_i(x_{i-1})$ has a unique *k*-fold equilibrium in **I'**. Combining both results we conclude that from any starting point $x_0$ in **R** the cyclical sequence defined by $x_i := A_i(x_{i-1})$ has a unique *k*-fold equilibrium in **I** ∩ **I'** = [z',z].

If all $Q_i$ are flat at zero then the equilibrium is only unique with respect to **I**. Note that in that case each point in the equilibrium is an intersection point of some non-trivial yield curve $P_i$ and the *x*-axis.

THEOREM IV.7    (zero sum of yields)    Let $(S_i, T_i)_{i \in \mathbf{N}}$ be a basic instance of the standard model. Let P and Q be entities in $(S_i, T_i)_{i \in \mathbf{N}}$. Let the balance sequence $(u_i)_{i \in \mathbf{N}}$ be a *k*-fold equilibrium of $(S_i, T_i)_{i \in \mathbf{N}}$ with respect to P and Q (see definition III.9). Then we have:

$$\sum_{1 \leq i \leq k} P_i(u_i) - Q_i(-u_i) = 0.$$

PROOF Note that by construction we have $u_{i+1} = A_i(u_i) = u_i + P_i(u_i) - Q_i(-u_i)$. Further, we have $u_{i+k} = u_i$, since $(u_i)_{i \in \mathbf{N}}$ is a *k*-fold equilibrium and hence cyclical. Now the claim follows by induction.                                                                                                      □

The theorem shows that any equilibrium constitutes a sequence where the total some of yields exchanged between P and Q is zero.

### VI. REFERENCES
[1]    W.P. Weijland, *Mathematical Foundations for the Economy of Giving*, ArXiv Categories: q-fin.GN, Report 1401.4664, 2014.

[2]    R.M. Brooks a& K. Schmitt, *The Contraction Mapping Principle and Some Applications*, Electronic Journal of Differential Equations, Monograph 09, ISSN: 1072-6691, 2009.

[3]  J.C.M. Baeten & W.P. Weijland, *Process Algebra* (248p), Cambridge University Press, 1990.

[4]  A. Karadimas, *Relating Economy of Giving to Peer-to-Peer File Sharing Technology*, University of Amsterdam, 2014.

[5]  N. Kiyotaki & R. Wright, *On Money as a Medium of Exchange, Journal of Political Economy*, vol. 97, no. 41 K 1989, The University of Chicago, 1989.

[6]  A. Oberhauser, *Decentralized Public Ledger as Enabler for the Gift Economy at Scale*, VU University Amsterdam, 2014.

[7]  J.A. Bergstra and W.P. Weijland, *Bitcoin: a Money-like Informational Commodity*, ArXiv categories: cs.CY, Report 1402.4778, 2014.